\documentclass[aps,pre,reprint,superscriptaddress,floatfix]{revtex4-1}

\usepackage{graphicx}

\usepackage{amsmath}  
\usepackage{esint} 
\usepackage{color} 

\usepackage{dcolumn}
\usepackage{bm}

\usepackage[normalem]{ulem}
\useunder{\uline}{\ul}{}

\usepackage{lipsum}

\begin{document}

\title{Hierarchical burst model for complex bursty dynamics}

\author{Byoung-Hwa Lee}
\affiliation{Department of Physics, Pohang University of Science and Technology, Pohang 37673, Republic of Korea}
\affiliation{Asia Pacific Center for Theoretical Physics, Pohang 37673, Republic of Korea}
\author{Woo-Sung Jung}
\email{wsjung@postech.ac.kr}
\affiliation{Department of Physics, Pohang University of Science and Technology, Pohang 37673, Republic of Korea}
\affiliation{Department of Industrial and Management Engineering, Pohang University of Science and Technology, Pohang 37673, Republic of Korea}
\affiliation{Asia Pacific Center for Theoretical Physics, Pohang 37673, Republic of Korea}
\author{Hang-Hyun Jo}
\email{hang-hyun.jo@apctp.org}
\affiliation{Asia Pacific Center for Theoretical Physics, Pohang 37673, Republic of Korea}
\affiliation{Department of Physics, Pohang University of Science and Technology, Pohang 37673, Republic of Korea}
\affiliation{Department of Computer Science, Aalto University, Espoo FI-00076, Finland}

\date{\today}

\begin{abstract}
    Temporal inhomogeneities observed in various natural and social phenomena have often been characterized in terms of scaling behaviors in the autocorrelation function with a decaying exponent $\gamma$, the interevent time distribution with a power-law exponent $\alpha$, and the burst size distributions. Here the interevent time is defined as a time interval between two consecutive events in the event sequence, and the burst size denotes the number of events in a bursty train detected for a given time window. In order to understand such temporal scaling behaviors implying a hierarchical temporal structure, we devise a hierarchical burst model by assuming that each observed event might be a consequence of the multi-level causal or decision-making process. By studying our model analytically and numerically, we confirm the scaling relation $\alpha+\gamma=2$, established for the uncorrelated interevent times, despite of the existence of correlations between interevent times. Such correlations between interevent times are supported by the stretched exponential burst size distributions, for which we provide an analytic argument. In addition, by imposing conditions for the ordering of events, we observe an additional feature of log-periodic behavior in the autocorrelation function. Our modeling approach for the hierarchical temporal structure can help us better understand the underlying mechanisms behind complex bursty dynamics showing temporal scaling behaviors.
\end{abstract}

\maketitle

\section{Introduction}

Events in temporal patterns of natural and social phenomena have often been found to be inhomogeneously distributed in time. Examples include solar flares~\cite{Wheatland1998Waitingtime}, earthquakes~\cite{Corral2004Longterm, deArcangelis2006Universality}, neuronal firing~\cite{Kemuriyama2010Powerlaw}, and human social activities~\cite{Barabasi2005Origin, Karsai2018Bursty}. Such temporal inhomogeneities in event sequences have been studied in terms of bursts, which are rapidly occurring events in short-time periods, alternating with long periods of inactivity. It is known that many dynamical processes, such as spreading or diffusion, taking place in a network of individuals are strongly influenced by the bursty temporal patterns of individuals and/or interactions between them~\cite{Vazquez2007Impact, Karsai2011Small, Miritello2011Dynamical, Rocha2011Simulated, Holme2012Temporal, Jo2014Analytically, Delvenne2015Diffusion}. Therefore, it is of utmost importance to comprehensively characterize temporal inhomogeneities, not only for understanding various complex dynamics but also for predicting and even controlling them, if possible.

In order to characterize the temporal inhomogeneities in event sequences, we first denote the event sequence by $x(t)$ that has a value of $1$ at the moment of event occurred, $0$ otherwise. Then one can measure an autocorrelation function with delay time $t_d$ as
\begin{equation}
    \label{eq:autocorrel}
    A(t_d)=\frac{\langle x(t)x(t+t_{d})\rangle_{t}-\langle x(t)\rangle_{t}^{2}}{\langle x(t)^{2}\rangle_{t}-\langle x(t)\rangle_{t}^{2}},
\end{equation}
where $\langle \cdot \rangle_{t}$ is a time average. For event sequences with long-range memory effects, the autocorrelation function often shows a power-law decaying behavior as 
\begin{equation}
    \label{eq:auto_gamma}
    A(t_{d})\sim t_{d}^{-\gamma}
\end{equation}
with a decaying exponent $\gamma$. In general, temporal correlations characterized by $A(t_d)$ can be understood in terms of (i) interevent times and (ii) correlations between interevent times~\cite{Jo2017Modeling}. Here the interevent time is defined as a time interval between two consecutive events, denoted by $\tau$. The statistics of interevent times have been described by the interevent time distribution, while the correlations between interevent times have been studied in terms of burst size distributions~\cite{Karsai2012Universal, Karsai2018Bursty}.

In many empirical datasets showing temporal inhomogeneities, the interevent time distribution $P(\tau)$ has been characterized by a power-law function as
\begin{equation}
    \label{eq:iet_alpha}
    P(\tau) \sim \tau^{-\alpha}
\end{equation}
with $\alpha$ denoting the power-law exponent~\cite{Karsai2018Bursty}. It has been proved that when interevent times are fully uncorrelated with each other, the power-law exponent $\alpha$ of the interevent time distribution is related to the decaying exponent $\gamma$ of the autocorrelation function such that $\alpha+\gamma=2$ for $1< \alpha <2$~\cite{Lowen1993Fractal, Vajna2013Modelling}. On the other hand, the correlations between interevent times have been studied in terms of bursty trains~\cite{Karsai2012Universal}. A bursty train or burst is defined as a set of events such that interevent times between any two consecutive events in the same burst are less than or equal to a given time window $\Delta t$, while those between events in different bursts are larger than $\Delta t$. We denote the number of events in each burst by $b$, and its distribution by $Q_{\Delta t}(b)$. Since the uncorrelated interevent times result in the exponential function of burst size distributions, any deviation from the exponential function may indicate the existence of correlations between interevent times, often called correlated bursts~\cite{Karsai2012Universal, Karsai2012Correlated, Wang2015Temporal, Jo2015Correlated, Jo2017Modeling, Jo2018Limits}. In particular, the power-law burst size distributions for a wide range of $\Delta t$ have been observed in earthquakes, neuronal activities, and human communication patterns~\cite{Karsai2012Universal}, i.e., 
\begin{equation}
    \label{eq:burst_beta}
    Q_{\Delta t}(b) \sim b^{-\beta}
\end{equation}
with $\beta$ denoting the power-law exponent. Here one can ask a question about how strong correlations between interevent times should be present to violate the scaling relation $\alpha+\gamma=2$ derived for the uncorrelated case. Our understanding on this issue is far from complete, except for few recent works~\cite{Rybski2012Communication, Vajna2013Modelling, Jo2017Modeling}.

Along with various characterization methods for the bursty temporal patterns, a number of modeling approaches have been suggested to understand the underlying mechanisms behind such temporal inhomogeneities~\cite{Karsai2018Bursty}. In the case with human dynamics, we find several modeling approaches, such as priority queuing models~\cite{Barabasi2005Origin, Vazquez2006Modeling}, inhomogeneous Poissonian processes~\cite{Malmgren2008Poissonian, Malmgren2009Universality}, self-exciting point processes~\cite{Masuda2013Selfexciting, Jo2015Correlated}, and reinforcement models~\cite{Karsai2012Universal, Wang2014Modeling}. Although previous modeling approaches have been successful for understanding the observed temporal inhomogeneities to some extent, we here take an alternative modeling approach, inspired by the scaling behaviors in Eqs.~\eqref{eq:auto_gamma}--\eqref{eq:burst_beta}, indicating a hierarchical temporal structure in various complex systems. In order to understand the hierarchical temporal structure, we devise a hierarchical burst model by assuming that each observed event in an event sequence might be a consequence of the multi-level causal or decision-making process: A seed event at the zeroth level induces other events at the first level, each of which in turn leads to other events at the second level, and so on. Thanks to the simplicity of our model, we can derive a fractal dimension $d_f$ of the event sequence, the decaying exponent $\gamma$ of the autocorrelation function, and the power-law exponent $\alpha$ of the interevent time distribution to confirm the scaling relation $\alpha+\gamma=2$, while the derivation of the burst size distributions turns out to be not straightforward. Our modeling approach can help us better understand the underlying mechanisms behind complex bursty dynamics, e.g., in terms of a hierarchical task organization for human dynamics. We also note that hierarchical document streams have been modeled using an infinite-state automaton~\cite{Kleinberg2002Bursty}, sharing the goal with our approach.

Our paper is organized as follows: In Sec.~\ref{sec:model}, after introducing the hierarchical burst model, we study our model analytically and numerically in terms of the scaling behaviors of the fractal temporal structure, the autocorrelation function, and the interevent time distribution. Then we numerically obtain the stretched exponential burst size distributions, for which we provide an analytic argument. We also discuss the effect of imposing the ordering of events in terms of log-periodicity in the autocorrelation function. Finally, we conclude our work in Sec.~\ref{sec:conclusion}.

\begin{figure}[!t]
    \includegraphics[width=\columnwidth]{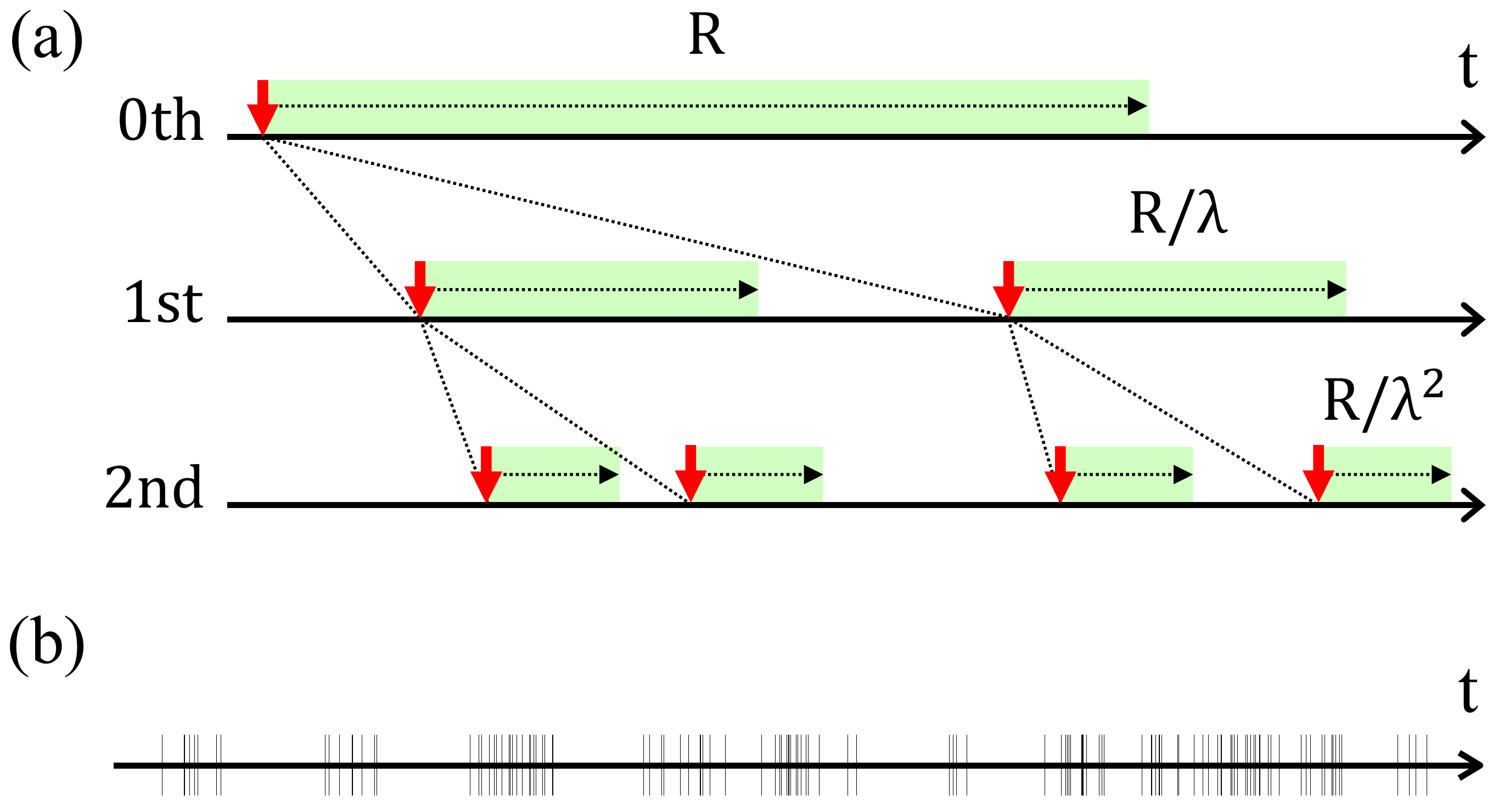}\centering
    \caption{(a) Schematic diagram of the hierarchical burst model with $\eta=2$ up to the second level. Red vertical arrows indicate the events at each level, each of which is assigned an induction interval (light green shade with horizontal dotted arrow) for events at the next level. (b) An example of the event sequence generated by the model with $\eta=2$, $\lambda=2.5$, $R=1$, and $L=8$. See the text for the details of the model.}
    \label{Fig1}
\end{figure}

\section{Model and results}\label{sec:model}

\subsection{Model definition}\label{subsec:def}

We introduce the hierarchical burst model by assuming that each observed event in an event sequence might be a consequence of the multi-level causal or decision-making process: A seed event at the zeroth level induces other events at the first level, each of which in turn leads to other events at the second level, and so on. Then the events at the final level compose the event sequence, while events at other levels are considered to be unobservable or hidden~\footnote{It is also possible to consider events at the intermediate levels observable, which is found to lead to the same conclusions (not shown).}. Precisely, for the levels of $l=0,1,\cdots, L-1$, each event at the $l$th level induces exactly $\eta$ events at the $(l+1)$th level, where $\eta>1$. If the timing of one event at the $l$th level is denoted by $t_l$, the events induced by that event take place uniformly at random in the time interval of $(t_l,t_l + R/\lambda^l]$, which we call an induction interval. Here $R$ is the induction interval assigned to the seed event, and $\lambda>1$ denotes the contraction factor between the induction intervals of consecutive levels. See Fig.~\ref{Fig1}(a) for the schematic diagram and Fig.~\ref{Fig1}(b) for an event sequence generated using $\eta=2$, $\lambda=2.5$, $R=1$, and $L=8$, resulting in $n=\eta^L=256$ events.

\begin{figure*}[!t]
    \includegraphics[width=\textwidth]{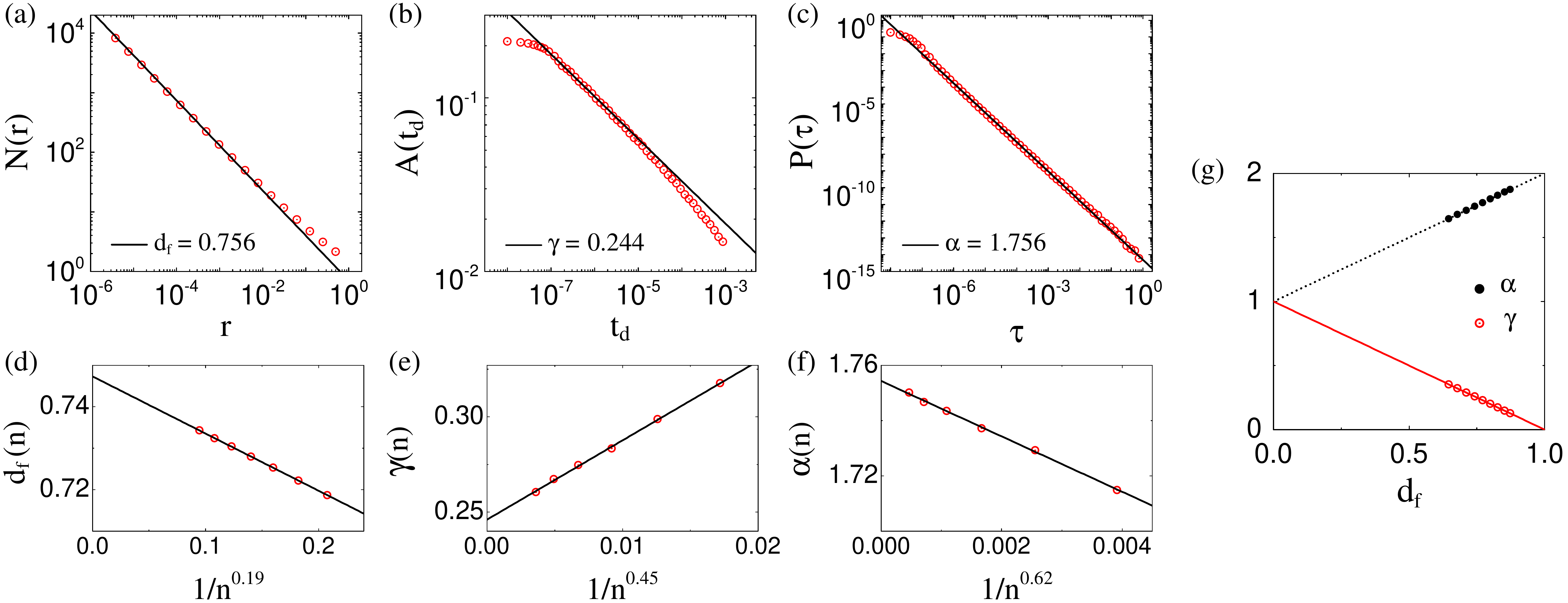}\centering
    \caption{Simulation results of the hierarchical burst model using $\eta=2$. In (a--c), the box counting result $N(r)$ for measuring the fractal dimension, the autocorrelation function $A(t_d)$, and the interevent time distribution $P(\tau)$ in the case with $\lambda=2.5$ and $L=18$ (circles) are compared to the analytical results for $d_f$, $\gamma$, and $\alpha$ (solid lines), respectively. Each curve was averaged over $500$ realizations of event sequences. In (d--f), for the same value of $\lambda=2.5$, we plot the fractal dimension $d_f(n)$, the decaying exponent $\gamma(n)$ of the autocorrelation function, and the power-law exponent $\alpha(n)$ of the interevent time distribution, as functions of $n=\eta^L$, i.e., the number of events in the event sequence (circles). For fitting the data, we adopt the functional form of $f(n)=f(\infty)+an^{-\nu}$ with $\nu>0$ (solid lines). In (g), we plot numerical values of $d_f$, $\gamma$, and $\alpha$ using various values of $\lambda$ for fixed $\eta=2$ and $L=18$, where each point (circle) was averaged over $200$ realizations of event sequences, to confirm the scaling relations of $\alpha=1+d_f$ (dotted line) and $\gamma=1-d_f$ (solid line).}
    \label{Fig2}
\end{figure*}

We focus on the case with one seed event, enabling us to set $R=1$ without loss of generality. Then the case with multiple seed events will be briefly discussed in Subsec.~\ref{subsec:multiSeed}. Since $\eta>1$ and $\lambda>1$, the induction interval decreases exponentially as a function of the level index $l$, while the number of events at the $l$th level is an exponentially increasing function of $l$. Hence, our model can be interpreted as a successive division of a big task (the seed event) into smaller tasks, ending up with the smallest unit of tasks (events at the final level) that are supposed to be executed in a bursty way. Our model can also be mapped to the one-dimensional Soneira-Peebles model~\footnote{Note that the original Soneira-Peebles (SP) model is isotropic as it was meant to describe the galaxy distributions in space, while our model is intrinsically directed due to the time asymmetry. Despite this difference, our model can be mapped to the one-dimensional SP model by shifting the timings of events at each level, but without affecting the statistical properties of the event sequence.}, which was originally introduced to generate self-similar galaxy distributions in two- or three-dimensional space~\cite{Soneira1978Computer, paredes1995clustering}, and recently applied to model a population landscape for human mobility~\cite{Hong2018Gravity}.

\subsection{Temporal scaling behaviors}\label{subsec:scaling}

Once an event sequence of $n=\eta^L$ events is generated by our model, we can derive its fractal dimension, autocorrelation function, and interevent time distribution, while the derivation of burst size distributions turns out to be not straightforward.

For calculating the fractal dimension $d_f$ of the event sequence, the box-counting method is used: We count the number of boxes of size $r$ needed to cover all events, which is denoted by $N(r)$. If $N(r)$ decays as a power law according to $r$, the corresponding power-law exponent defines the fractal dimension $d_f$, namely, 
\begin{equation}
    N(r)\sim r^{-d_f}.
\end{equation}
When the box size is given as $r=1/\lambda^l$ for $l=0,1,\cdots,L-1$, we get $N(r)\simeq \eta^l$, leading to 
\begin{equation}
    \label{eq:fractal}
    \eta=\lambda^{d_f}\ \textrm{or}\ d_{f}=\frac{\ln \eta}{\ln \lambda}.
\end{equation}
Here we have assumed that the boxes covering events or induction intervals at the same level do not necessarily overlap, or that even when they overlap, its effect would be negligible in estimating the fractal dimension. We will use this assumption for the following analysis, while its effect will be numerically studied in Subsec.~\ref{subsec:overlap}. Note that $d_f$ cannot be larger than the spatial (or temporal) dimension of $1$, even when $\eta>\lambda$. However, we will consider only the case with $\eta<\lambda$.

As evident in Eq.~\eqref{eq:autocorrel}, the autocorrelation function $A(t_d)$ with delay time $t_d$ essentially measures the possibility of finding two events observed in $t$ and $t+t_d$, no matter how many events occur between them. The number of events within the range of $t_d$ from any event is of the order of $t_d^{d_f}$ using Eq.~\eqref{eq:fractal}. Therefore, the number of events in the range of $(t_d,t_d+dt_d)$ is of the order of $t_d^{d_f-1}dt_d$, implying that the autocorrelation function has the form of
\begin{equation}
    \label{eq:gamma}
    A(t_{d}) \sim t_d^{-\gamma}\ \textrm{with}\ \gamma=1-d_f.
\end{equation}

Next, we derive the interevent time distribution $P(\tau)$. Let us consider $\eta^{l +1}$ events at the $(l+1)$th level. Among them, events induced by the same event in $t=t_l$ at the $l$th level will be found in the range of $(t_l,t_l + 1/\lambda^l]$. Thus, the interevent times between events induced by the same $l$th-level event must be of the order of $1/\lambda^l$. On the other hand, events induced by different $l$th-level events will be separated by interevent times larger than $1/\lambda^l$. The number of such cases corresponds to that of events at the $l$th level, i.e., $\sim \eta^l$. Hence one can write
\begin{equation}
    \Pr\left(\tau> \frac{1}{\lambda^l} \right) \sim \eta^l.
\end{equation}
By using $F(\tau)\equiv \int_\tau^\infty P(\tau')d\tau'$ and the relation $\eta=\lambda^{d_f}$ in Eq.~\eqref{eq:fractal}, one gets $F(\tau)\sim \tau^{-d_f}$, leading to
\begin{equation}
    \label{eq:alpha}
    P(\tau) \sim \tau^{-\alpha}\ \textrm{with}\ \alpha=1+d_f.
\end{equation}
Since $0\leq d_f\leq 1$ in our model, the value of $\alpha$ is limited to the range of $[1,2]$. 

Finally, combining the results in Eq.~\eqref{eq:gamma} and Eq.~\eqref{eq:alpha}, we obtain the scaling relation between $\alpha$ and $\gamma$ as follows:
\begin{equation}
    \label{eq:scaling}
    \alpha+\gamma=2,
\end{equation}
which turns out to hold irrespective of $d_f$, i.e., irrespective of $\eta$ and $\lambda$. This scaling relation has been derived for the case that interevent times are fully uncorrelated with each other~\cite{Lowen1993Fractal, Vajna2013Modelling}. Hence, this result in Eq.~\eqref{eq:scaling} may indicate that the correlations between interevent times in our model are not strong enough to violate the scaling relation in Eq.~\eqref{eq:scaling}. In order to tackle this issue, we will study the burst size distribution $Q_{\Delta t}(b)$ in Subsec.~\ref{subsec:burstSize}.

\begin{figure*}[!t]
    \includegraphics[width=\textwidth]{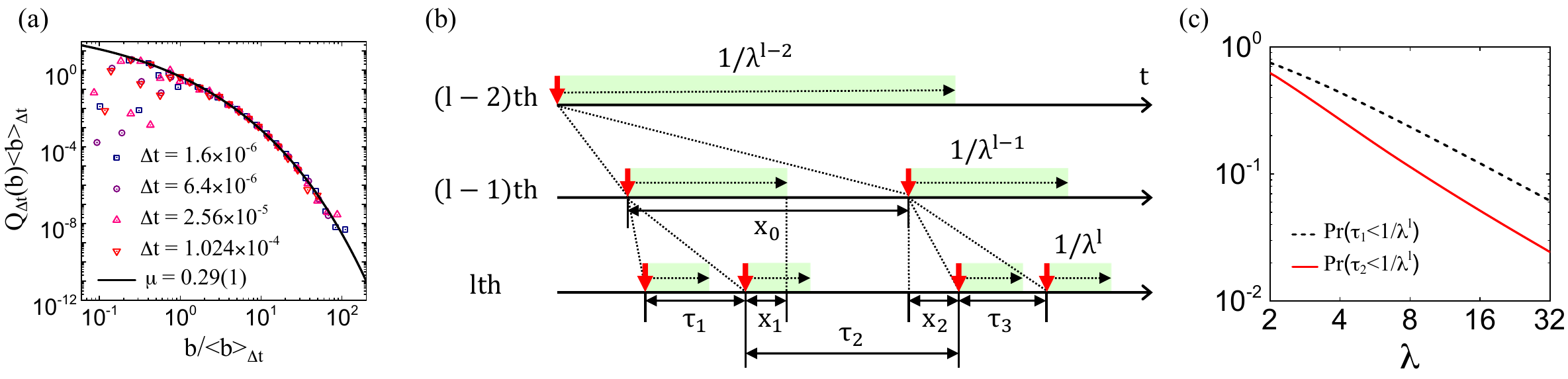}\centering
    \caption{(a) Simulation results of the burst size distributions $Q_{\Delta t}(b)$ for various time windows $\Delta t$ by the hierarchical burst model using $\eta=2$, $\lambda=2.5$, and $L=18$ (symbols), fitted with a stretched exponential function (black solid line). Here $\langle b\rangle_{\Delta t}$ is the average burst size for a given $\Delta t$. (b) A schematic diagram for the analytic calculation of probability distributions of interevent time between events induced by the same (or different) events at the $(l-1)$th level, denoted by $P(\tau_1)$ [$P(\tau_2)$]. (c) Comparison between $\Pr(\tau_1<\Delta t)$ in Eq.~\eqref{eq:same} and $\Pr(\tau_2<\Delta t)$ in Eq.~\eqref{eq:diff} for $\Delta t= 1/\lambda^l$.}
    \label{Fig3}
\end{figure*}

For the numerical simulations of our model, we begin with one seed event in $t=0$ at the zeroth level. Then $\eta$ events are uniformly distributed in the range of $(0,1]$ at the first level. Each of these induced events in turn induces $\eta$ events at the second level in the range of $(t_1,t_1+1/\lambda]$, with $t_1$ denoting the timing of one of events at the first level. This induction process is repeated until the $L$th level is reached, leaving us $n=\eta^L$ events. For the demonstration, we focus on the case with $\eta=2$ and $\lambda=2.5$, with which one expects $d_f\approx 0.756$ from Eq.~\eqref{eq:fractal}, consequently $\gamma\approx 0.244$ and $\alpha\approx 1.756$ from Eq.~\eqref{eq:gamma} and Eq.~\eqref{eq:alpha}, respectively.

We analyze the generated event sequences for various values of $L$. For example, Fig.~\ref{Fig2}(a--c) shows the numerical results of $N(r)$, $A(t_d)$, and $P(\tau)$, all averaged over $500$ event sequences using $L=18$. We find that the estimated values of corresponding power-law exponents, i.e., $d_f$, $\gamma$, and $\alpha$, are comparable with those expected from the analysis, but with some visible deviations in cases of $d_f$ and $\gamma$. These deviations could be due to the finite-size effects. In order to study such finite-size effects, we estimate the above power-law exponents for various sizes of $n$, equivalently, for various values of $L=12,\cdots,18$, as shown in Fig.~\ref{Fig2}(d--f). The size-dependent power-law exponents are denoted by $d_f(n)$, $\gamma(n)$, and $\alpha(n)$, respectively. Each of these exponents is fitted with a functional form of $f(n)=f(\infty) + an^{-\nu}$ with $\nu>0$, from which the value of $f(\infty)$ is obtained. We find that $d_f(\infty)=0.75(1)$, $\gamma(\infty)=0.25(1)$, and $\alpha(\infty)=1.75(1)$, all consistent with those expected from the analysis within error bars. Based on these results, we conclude that the effects of overlapping induction intervals at the same level turn out to be negligible to the scaling relations. Finally, we numerically confirm the scaling relations in Eqs.~\eqref{eq:gamma} and~\eqref{eq:alpha} for various values of $\lambda$ when $\eta=2$ is fixed, as depicted in Fig.~\ref{Fig2}(g).

\subsection{Burst size distributions}\label{subsec:burstSize}

In order to scrutinize the existence of correlations between interevent times, we measure the burst size distributions $Q_{\Delta t}(b)$ for various values of the time window $\Delta t$. For example, the numerical results for $L=18$ are shown in Fig.~\ref{Fig3}(a). The curves of $Q_{\Delta t}(b)$ for a wide range of $\Delta t$, when properly normalized, turn out to collapse into the same curve for the range of $b>\langle b\rangle_{\Delta t}$, where $\langle b\rangle_{\Delta t}$ is the average burst size when the time window is given as $\Delta t$. This curve is well described by the stretched exponential function, i.e., 
\begin{equation}
    \label{eq:burstsize}
    Q_{\Delta t}(b) \sim \exp\left(-c_{\Delta t}b^\mu\right),
\end{equation}
with $\mu \approx 0.29(1)$ and $c_{\Delta t}$ denoting a proper coefficient depending on $\Delta t$. The fact that $Q_{\Delta t}(b)$ deviates from the exponential function indicates the existence of correlations between interevent times. At the same time, such correlations depicted in terms of stretched exponential functions might not be strong enough to violate the scaling relation between $\alpha$ and $\gamma$ in Eq.~\eqref{eq:scaling}. This conclusion is indeed consistent with the observations in Ref.~\cite{Jo2017Modeling}, in which even the power-law burst size distribution in the form of $Q_{\Delta t}(b)\propto b^{-\beta}$ could not violate the relation of $\alpha+\gamma=2$ unless it has a sufficiently heavy tail with $\beta<3$.

In order to understand why burst size distributions observed in our model are better described by a stretched exponential function rather than a power-law function, we study how likely it is to cluster events induced by the different events to the same burst for a given time window. The more likely such case happens, the burst size distribution can have a heavier tail. Precisely, for a given time window $\Delta t$, we calculate the probability of events induced by the different events to be clustered to the same burst, which is then compared to the probability of events induced by the same event to be clustered to the same burst.

For the analysis we consider the minimal case with $\eta=2$. When the time window is given as $\Delta t=1/\lambda^l$, we only need to consider the events at the $l$th, $(l-1)$th, and $(l-2)$th levels, as depicted in Fig.~\ref{Fig3}(b). The timescales at other levels are either too large or too small to be relevant to the analysis. We denote the timing of one event at the $(l-2)$th level by $t_{l-2}$. This event induces two events at the $(l-1)$th level, whose timings are respectively $t_{l-1,0}$ and $t_{l-1,1}$, satisfying
\begin{equation}
    t_{l-2}< t_{l-1,0}<t_{l-1,1}\leq t_{l-2}+\frac{1}{\lambda^{l-2}}.
\end{equation}
These two events at the $(l-1)$th level induce four events at the $l$th level, whose timings are respectively $t_{l,0}$, $t_{l,1}$, $t_{l,2}$, and $t_{l,3}$, satisfying
\begin{eqnarray}
    && t_{l-1,0}< t_{l,0}<t_{l,1}\leq t_{l-1,0}+\frac{1}{\lambda^{l-1}},\\
    && t_{l-1,1}< t_{l,2}<t_{l,3}\leq t_{l-1,1}+\frac{1}{\lambda^{l-1}}.
\end{eqnarray}
That is, the events in $t_{l,0}$ and $t_{l,1}$ are induced by the event in $t_{l-1,0}$, while the events in $t_{l,2}$ and $t_{l,3}$ are induced by the event in $t_{l-1,1}$. By assuming that $t_{l,1}<t_{l,2}$, we have three interevent times between events at the $l$th level:
\begin{eqnarray}
    \tau_1 &\equiv& t_{l,1}-t_{l,0},\\
    \label{eq:tau2}
    \tau_2 &\equiv& t_{l,2}-t_{l,1},\\
    \tau_3 &\equiv& t_{l,3}-t_{l,2},
\end{eqnarray}
see Fig.~\ref{Fig3}(b). Using the order statistics~\cite{David2003Order, Kivela2012Multiscale, Kim2016Measuring}, we get the distribution of $\tau_1$ as
\begin{equation}
    P(\tau_1)=2 \lambda^{l-1} (1-\lambda^{l-1}\tau_1)
\end{equation}
for $0<\tau_1\leq 1/\lambda^{l-1}$, which is the same as $P(\tau_3)$. Then the probability of clustering two events induced by the same $(l -1)$th-level event for a given $\Delta t=1/\lambda^l$ is calculated as
\begin{equation}
    \label{eq:same}
    \Pr \left( \tau_1<\frac{1}{\lambda^{l}} \right) = \Pr \left( \tau_3<\frac{1}{\lambda^{l}} \right) = \frac{2}{\lambda}-\frac{1}{\lambda^{2}}.
\end{equation}

Next, in order to derive the distribution of $\tau_2$, we rewrite $\tau_2$ in Eq.~\eqref{eq:tau2} as
\begin{equation}
    \tau_2=x_0+x_1+x_2-\frac{1}{\lambda^{l-1}},
\end{equation}
where 
\begin{eqnarray}
    x_0 &\equiv& t_{l-1,1}-t_{l-1,0},\\
    x_1 &\equiv& t_{l-1,0}+\frac{1}{\lambda^{l-1}}-t_{l,1},\\
    x_2 &\equiv& t_{l,2}-t_{l-1,1},
\end{eqnarray}
see Fig.~\ref{Fig3}(b). Here $x_0$ is indeed the interevent time between events at the $(l-1)$th level, leading to its distribution as
\begin{equation}
    \label{eq:tau0pdf}
    P(x_0)=2 \lambda^{l-2} (1-\lambda^{l-2}x_0)
\end{equation}
for $0<x_0\leq 1/\lambda^{l-2}$. We also get the distribution of $x_1$ as
\begin{equation}
    \label{eq:x1pdf}
    P(x_1)=2 \lambda^{l-1} (1-\lambda^{l-1}x_1)
\end{equation}
for $0<x_1\leq 1/\lambda^{l-1}$, which is the same as $P(x_2)$. We now calculate $\Pr(\tau_2<\Delta t)$, i.e.,
\begin{equation}
    \Pr \left( \tau_2<\frac{1}{\lambda^{l}} \right) = \Pr \left(x_0+x_1+x_2< \frac{1}{\lambda^{l}} +\frac{1}{\lambda^{l-1}} \right).
\end{equation}
As $x_i$s for $i=0,1,2$ are statistically independent of each other, using $t\equiv 1/\lambda^l +1/\lambda^{l-1}$ we rewrite the above equation as
\begin{equation}
    h(t)\equiv \Pr \left( \sum_{i=0}^2 x_i < t\right) = \prod_{i=0}^2 \int dx_i P(x_i) \theta\left(t- \sum_{i=0}^2 x_i \right),
\end{equation}
where $\theta(\cdot)$ is the Heaviside step function. Taking the Laplace transform, one gets
\begin{equation}
    \tilde h(s)=\frac{1}{s} \prod_{i=0}^2 \left[\int dx_i P(x_i)e^{-sx_i} \right].
\end{equation}
By plugging the Laplace transforms of $P(x_i)$ in Eqs.~\eqref{eq:tau0pdf} and~\eqref{eq:x1pdf} into the above equation, and then taking the inverse Laplace transform of $\tilde h(s)$, one can get $h(t)$. We finally obtain
\begin{eqnarray}
    \label{eq:diff}
    &&\Pr \left( \tau_2<\frac{1}{\lambda^{l}} \right) = \\
    &&\frac{11}{15\lambda} + \frac{131}{90\lambda^{2}} - \frac{1}{15\lambda^{3}} - \frac{3}{2\lambda^{4}} - \frac{2}{9\lambda^{5}} + \frac{1}{10\lambda^{6}} - \frac{1}{15\lambda^{7}} + \frac{1}{90\lambda^{8}}\nonumber
\end{eqnarray}
for the entire range of $\lambda>\eta=2$. By comparing Eq.~\eqref{eq:same} with Eq.~\eqref{eq:diff}, we conclude that for $\lambda>2$
\begin{equation}
    \label{eq:prcom}
    \Pr \left( \tau_2<\frac{1}{\lambda^{l}} \right) < \Pr \left( \tau_1<\frac{1}{\lambda^{l}} \right),
\end{equation}
as numerically shown in Fig.~\ref{Fig3}(c). This inequality holds for any level index $l$, implying that the chance of clustering events induced by the different events at the previous level must be low at any level. This low chance in turn lowers the possibility of finding big bursts, giving us a hint at the reason why burst size distributions in our model do not show a heavier tail than the stretched exponential function. 

\begin{figure}[!t]
    \includegraphics[width=\columnwidth]{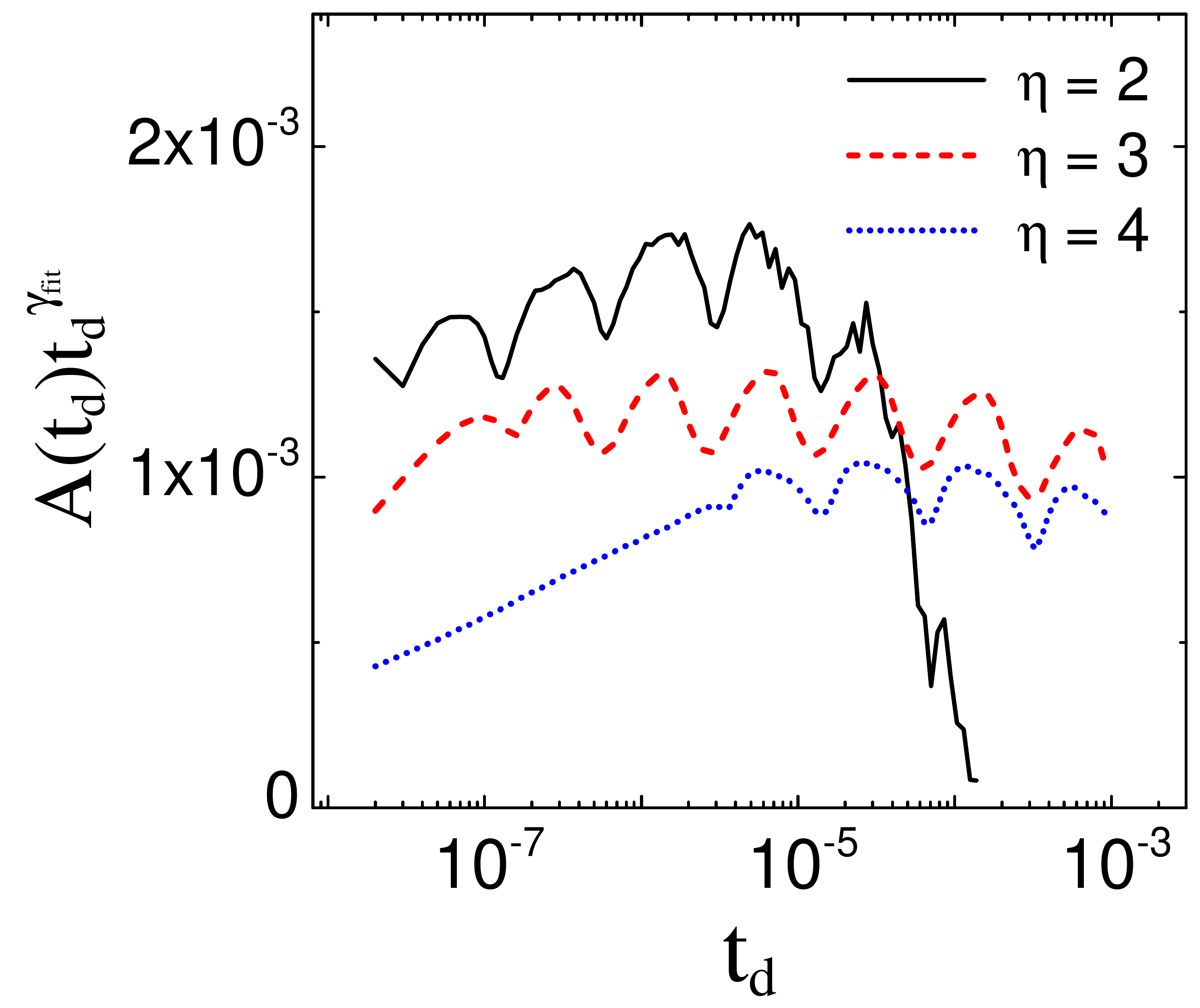}\centering
    \caption{Log-periodic behaviors in the autocorrelation functions of the hierarchical burst model using several values of $(\eta,L)=(2,17)$, $(3,11)$, and $(4,9)$ for a fixed $\lambda=4.8$ under conditions in Eqs.~\eqref{eq:non1} and~\eqref{eq:non2} for non-overlapping induction intervals, where all curves were averaged over $200$ realizations of event sequences. For each curve, we have used the best fit value of $\gamma_{\rm fit}$ for the best presentation of the log-periodicity. The curve for $\eta=2$ was vertically shifted for the clear presentation.}
    \label{Fig4}
\end{figure}

\subsection{Effect of non-overlapping induction intervals}\label{subsec:overlap}

Our model allows induction intervals at the same level to overlap with each other, although its effects turn out to be irrelevant to the scaling relations between $d_f$, $\gamma$, and $\alpha$, as discussed in Subsec.~\ref{subsec:scaling}. Let us consider two events at the $(l-1)$th level, which respectively take place in times $t_{l-1,0}$ and $t_{l-1,1}$, with $t_{l-1,0}<t_{l-1,1}$. Since induction intervals can overlap, it is possible that some events induced by the event in $t_{l-1,0}$ take place later than other events induced by the event in $t_{l-1,1}$, e.g., in the case when $t_{l,0}<t_{l,2}<t_{l,1}<t_{l,3}$ in Subsec.~\ref{subsec:burstSize}. This situation can be called an event crossing. The occurrence of event crossing may cause some problems, e.g., in the context of task executions: Although the tasks are supposed to be executed sequentially, they can be executed out of order, if possible. In order to avoid such event crossings, we impose a rule for the non-overlapping induction intervals. When one event in $t_{l-1}$ at the $(l-1)$th level induces $\eta$ events at the $l$th level, their timings, denoted by $t_{l,i}$ or $t_{l,j}$ for $i,j=1,\cdots,\eta$, are to satisfy the following conditions at each level:
\begin{eqnarray}
    \label{eq:non1}
    && t_{l,i} \in \left(t_{l-1},t_{l-1}+\frac{1}{\lambda^{l-1}}-\frac{1}{\lambda^{l}}\right],\\
    \label{eq:non2}
    && \left|t_{l,i}- t_{l,j}\right| \geq \frac{1}{\lambda^{l}}\ \textrm{for}\ i\neq j.
\end{eqnarray}
By the first condition in Eq.~\eqref{eq:non1} any descendent events of the event in $t_{l-1}$ are forced to remain in the range of $(t_{l-1},t_{l-1}+1/\lambda^{l-1}]$. The second condition in Eq.~\eqref{eq:non2} prohibits the induction intervals at the same level from overlapping with each other.

\begin{figure}[!t]
    \includegraphics[width=\columnwidth]{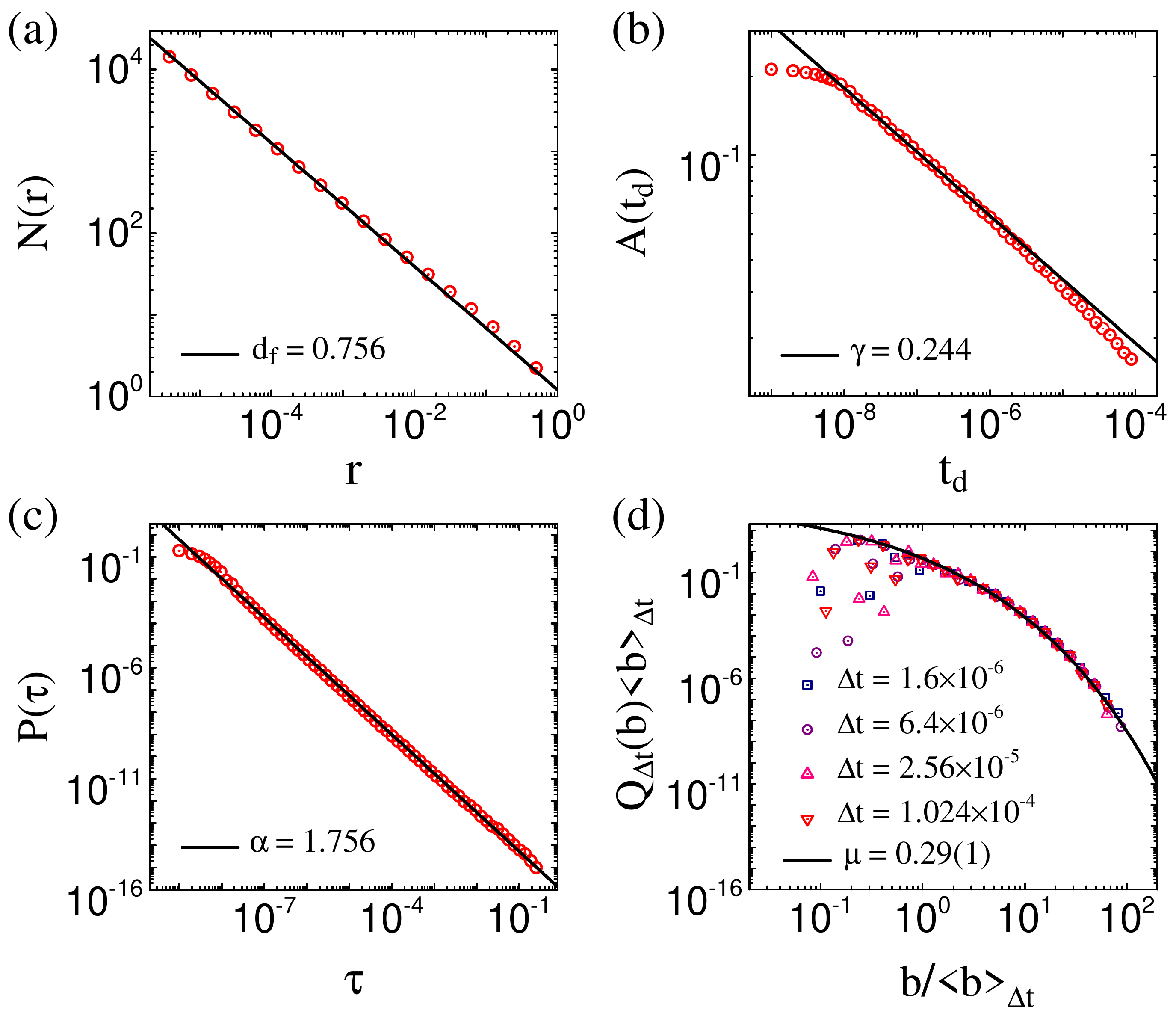}\centering
    \caption{Simulation results of the hierarchical burst model using $\eta=2$, $\lambda=2.5$, $R=0.1$, and $L=18$, with $10$ seed events at the zeroth level: (a) Box counting result for measuring the fractal dimension, (b) the autocorrelation function, (c) the interevent time distribution, and (d) the burst size distributions for various values of $\Delta t$. Each curve was averaged over $100$ realizations of event sequences. Black solid lines in (a--c) represent the analytic results, while the black solid line in (d) shows a stretched exponential function fitted to the data.}
    \label{Fig5}
\end{figure}

By performing numerical simulations, we find that the fractal dimension, the interevent time distribution, and the burst size distributions show overall the same behaviors as in the original version of our model (not shown). However, the autocorrelation function shows a qualitatively different behavior, as shown in Fig.~\ref{Fig4}. We find a power-law decaying function coupled with log-periodic behavior, say,
\begin{equation}
    A(t_d)\sim t_d^{-\gamma}\left[c_0+\cos(\omega \ln t_d+\phi)\right]
\end{equation}
with appropriate constants $c_0$, $\omega$, and $\phi$. In particular, we can relate the frequency $\omega$ to the contraction factor $\lambda$, based on the observation that the distance between consecutive peak times of $A(t_d)$ increases by a factor of $\lambda$ mainly due to Eq.~\eqref{eq:non2}. Precisely, let us consider a simple log-periodic function of $g(t)=\cos(\omega\ln t)$. The peak times are determined by $g(t_k)=1$, i.e., $t_k=e^{2\pi k/\omega}$ for integers $k$. As the distance between the $k$th and $(k+1)$th peak times is larger than the distance between the $(k-1)$th and $k$th peak times by a factor $\lambda$, one can write 
\begin{equation}
    t_{k+1}-t_k=\lambda (t_k-t_{k-1}),
\end{equation}
leading to the relation between $\omega$ and $\lambda$ as follows:
\begin{equation}
    \label{eq:omega}
    \omega =\frac{2\pi}{\ln \lambda}.
\end{equation}
For example, when $\lambda=4.8$, we get $\omega\approx 4.006$ from Eq.~\eqref{eq:omega}, which is comparable with the numercial value of $\omega=4.02(3)$ estimated from the curve for $\eta=3$ in Fig.~\ref{Fig4}. We also find that $\omega$ is not a function of $\eta$, which is probably because the condition in Eq.~\eqref{eq:non2} can be imposed irrespective of $\eta$.

\subsection{Case with multiple seed events}\label{subsec:multiSeed}

So far we have considered the case only with one seed event at the zeroth level. Here we test if our conclusions in the case with a single seed event are robust with respect to the number of seed events at the zeroth level. For this, we perform the numerical simulations of our original model, i.e., allowing induction intervals to overlap, with $10$ seed events whose timings are randomly chosen in the range of $[0,1]$. In this case, we set the induction interval for each seed event as $R=0.1$. The numerical results are summarized in Fig.~\ref{Fig5}, showing overall the same scaling behaviors of the fractal dimension, the autocorrelation function, and the interevent time distribution. We also find the stretched exponential function with the same value of $\mu=0.29(1)$ in Eq.~\eqref{eq:burstsize} fitted well to the burst size distributions.

\section{Conclusion}\label{sec:conclusion}

We have studied the hierarchical burst model for the hierarchical temporal structure by assuming that an observed event sequence is generated by a multi-level causal or decision-making process. A seed event at the zeroth level induces $\eta$ events at the first level, each of which in turn induces other $\eta$ events at the second level, and so on. The interval for the induction is assumed to decrease by a contraction factor $\lambda$ from one level to the next level. Only the events at the final level are considered to be observed in the event sequence. We first analyze the model by deriving the analytic solutions for the fractal dimension $d_f=\ln\eta/\ln\lambda$, the autocorrelation function with power-law exponent $\gamma=1-d_f$, and the interevent time distribution with power-law exponent $\alpha=1+d_f$. We immediately obtain $\alpha+\gamma=2$, irrespective of $d_f$. This scaling relation has been derived for the case that interevent times are fully uncorrelated with each other~\cite{Lowen1993Fractal, Vajna2013Modelling}. The scaling relations of $\gamma=1-d_f$ and $\alpha=1+d_f$ have also been numerically confirmed.

On the other hand, it turns out that the burst size distributions are not straightforward to analyze. By performing numerical simulations, we find the stretched exponential function for the burst size distributions, implying the existence of correlations between interevent times, often called correlated bursts. However, such correlations are not strong enough to violate the scaling relation $\alpha+\gamma=2$. For the stretched exponential burst size distributions, we provide an argument based on an analytical calculation. We also find that by imposing non-overlapping induction intervals for the ordering of events, the autocorrelation function is described by a power-law decaying function coupled with log-periodic behavior, whose frequency is related to the contraction factor $\lambda$.

Despite the debate on the functional form of burst size distributions~\cite{Karsai2012Universal, Jiang2016Twostate}, one can extend our model to reproduce the power-law burst size distributions as evident in some empirical data analysis~\cite{Karsai2012Universal}, which then can help us to understand the correlations between interevent times in the context of the hierarchical burst structure~\cite{Jo2017Modeling}. We also remark that $1\leq \alpha\leq 2$ in our model, while its empirical values are often found to be out of the range of $[1,2]$, as summarized in Ref.~\cite{Karsai2018Bursty}. This requires us to devise more flexible hierarchical burst models showing a wide range of $\alpha$. Finally, our model can be extended to incorporate a number of complex realistic situations. For example, we can consider the context of events~\cite{Jo2013Contextual} and a network of interacting individuals~\cite{Albert2002Statistical, Boccaletti2006Complex, Newman2010Networks, Holme2012Temporal}, whose interaction activities are described by complex bursty dynamics.

\begin{acknowledgments}
W.-S.J. was supported by Basic Science Research Program through the National Research Foundation of Korea (NRF) funded by the Ministry of Education (2016R1D1A1B03932590). H.-H.J. was supported by Basic Science Research Program through the National Research Foundation of Korea (NRF) funded by the Ministry of Education (2015R1D1A1A01058958).
\end{acknowledgments}

\bibliographystyle{apsrev4-1}
%


\end{document}